# Prerequisites for International Exchanges of Health Information: Comparison of Australian, Austrian, Finnish, Swiss, and US Privacy Policies

Adj/Prof Hanna Suominen
Research School of Computer Science (RSCS),The Australian National University (ANU); Data61, Commonwealth Scientific and Industrial Research Organisation (CSIRO); Faculty of Education, Science, Technology and Maths, University of Canberra (UC); and Department of Information Technology, University of Turku (UTU)
ANU, Building 108 (CSIT)
Acton 2601 ACT, AUSTRALIA
+61 2 612 532 57
hanna.suominen@anu.edu.au

Authors' Contributions: H Suominen and L Ohno-Machado were responsible for the study conception and design. H Suominen drafted the paper and reviewed the Australian and Finnish situations. The Austrian, Finnish, Swiss, and US contents were reviewed by G Schreier, S Salanterä, H Müller, and L Ohno-Machado, respectively. All authors commented critically on the study.

Prof Henning Müller
Business Information Systems, HES-SO University of Applied Sciences Western Switzerland, Sierre, Switzerland; Medical Faculty, University of Geneva, Geneva, Switzerland; and Martinos Center for Biomedical Imaging, Harvard Medical School, Boston, MA, USA
henning.mueller@hevs.ch

Prof Lucila Ohno-Machado
Division of Biomedical Informatics, Department of Medicine, University of California, San Diego, CA, USA
machado@ucsd.edu

Prof Sanna Salanterä
Department of Nursing Science, UTU and Turku University Hospital, Turku, Finland
sansala@utu.fi

Prof Günter Schreier
Digital Safety and Security Department, AIT Austrian Institute of Technology, Graz, Austria
guenter.schreier@ait.ac.at

Adj/Prof Leif Hanlen
Data61, CSIRO; RSCS, ANU; and Faculty of Health, UC; Canberra, Australia
leif.hanlen@data61.csiro.au

## ABSTRACT
Capabilities to exchange health information are critical to accelerate discovery and its diffusion to healthcare practice. However, the same ethical and legal policies that protect privacy hinder these data exchanges, and the issues accumulate if moving data across geographical or organizational borders. This can be seen as one of the reasons why many health technologies and research findings are limited to very narrow domains. In this paper, we compare how using and disclosing personal data for research purposes is addressed in Australian, Austrian, Finnish, Swiss, and US policies with a focus on text data analytics. Our goal is to identify approaches and issues that enable or hinder international health information exchanges. As expected, the policies within each country are not as diverse as across countries. Most policies apply the principles of accountability and/or adequacy and are thereby fundamentally similar. Their following requirements create complications with re-using and re-disclosing data and even secondary data: 1) informing data subjects about the purposes of data collection and use, before the dataset is collected; 2) assurance that the subjects are no longer identifiable; and 3) destruction of data when the research activities are finished. Using storage and compute cloud services as well as other exchange technologies on the Internet without proper permissions is technically not allowed if the data are stored in another country. Both legislation and technologies are available as vehicles for overcoming these barriers. The resulting richness in information variety will contribute to the development and evaluation of new clinical hypotheses and technologies.

## 1. INTRODUCTION
*Health information* includes all health-related content from all specialties, organizations, regions, territories, and countries. Examples include private personal data on *electronic health records* (EHRs) as well as health sciences papers and other publicly-available information on the Internet. Their electronic recording enables data to become potentially accessible through *information and communications technologies* (ICT) for the purposes of improving health and healthcare. These *benefits* are typically reached through supporting situational awareness, decision-making, and knowledge discovery, as is illustrated by the increasing popularity of *text data analytics* [1-3].

However, this promise only holds if the potential benefits outweigh the *risks* [4-8]. A major risk of *compromising data subjects' privacy* is particularly evident in analyzing text (i.e., inabilities to be fully convinced that all privacy-sensitive information has been de-identified both in the explicit text and in between its lines) or big data (i.e., unforeseen possibilities to infer personal data after record linkages from multiple de-identified sources); health information may contain patients or healthcare workers' *personal data* such as identifiers, care details, and other sensitive matters.

Consequently, storing and using them requires careful compliance with *jurisdictional legislation*, *governance*, and *policies*, for *record/register research*. To summarize their timeline internationally, in 1940s, the *Nurnberg Code on for Human Experimentation* was published; in 1960s, the *Declaration of Helsinki on Ethical Principles for Medical Research Involving Human Subjects* was developed; in 1970s, some of the first laws for protecting *electronic data privacy* were established in *Europe* and the *Belmont Report for the Protection of Human Subjects of Biomedical and Behavioral Research* was created; in 1980s, the *Lisbon Declaration of the Rights of the Patient* was released and the *Organisation for Economic Co-operation and Development* (OECD) adopted their framework, which addresses not only personal data and privacy protection but also *movement of personal data across national borders*; in 1990s and 2000s, similar frameworks by the *European Commission* (EC) and *Asia-Pacific Economic Co-Operation Organisation* (APEC) followed together with the *US Personal Health Information Privacy Rule*; and since 2010s, EC has targeted individuals' protection with regard to the *globalization*, *new ICT*, and *processing of personal data* with specific sections on processing of personal data concerning health and processing for historical, statistical, and scientific research purposes [9]. The most widely used legal frameworks in EHR information exchanges are the EC and US frameworks from 1995 and 2002, respectively [10].

## 2. METHODS

We addressed this information exchange problem from an international perspective by comparing *Australian*, *Austrian*, *Finnish*, *Swiss*, and *US* frameworks by extending a paper on Australian and Finnish frameworks [9]. We chose these countries to cover America, Australia, EU (Austria and Finland members since 1995), and non-EU Europe (Switzerland).

We extracted the relevant legal frameworks from online databases of the legislative and other judicial information of Australia, Australian New South Wales (NSW), Austria, EU, Finland, Switzerland, Swiss Valais, the USA, and the US California, supplemented with information available on the websites of the jurisdictional ethics boards. We included both general frameworks for privacy protection and health specific ones.

For each country, we addressed the following research questions: What are the relevant legal frameworks? How can the process of gaining access to authentic health information be characterized in terms of ethics approvals, research permissions, and data subjects' informed consenting? How can data be exchanged, combined, and compared across jurisdictional borders or from a research project to another?

## 3. RESULTS

Even though international frameworks have an increasing role and harmonizing impact over the *national*, *territorial*, *regional*, and *organizational frameworks*, the requirements for data access and protection vary in states, territories, and cantons of the five countries (Table 1); each region significantly adds to the frameworks by establishing their own regulations and recommendations.

The *process of gaining access to authentic health information* for the purposes of scientific, historical, or statistical projects is typically straightforward, well guided, and has the following five steps: First, *preparations* take place and include developing a *research plan*, forming a *research group*, *naming its leader*, and writing an *ethics protocol* that encompasses at least the following aspects: studying the governance, policy, and legal frameworks; assuring that the proper permissions are furnished and legislation is followed; monitoring that the permissions cover all aspects of project; specifying the purposes of data collection, including justifications for the relevance of the data to the research plan and the amount of data to be collected; specifying data collection, storage, and protection which includes data access, destruction, use, modification, and disclosure; preparing user agreements; educating the data users on research ethics; answering to questions on research ethics; monitoring that good research practice is conformed; and intervening in problems.

Second, *permissions* are furnished: The study is *accepted by chief officers* of the jurisdiction. *Ethics approvals* are obtained from the approving authority, except in Austria and Finland, some projects are not subject to this. *Research permissions* are obtained at least from the each healthcare jurisdiction where data originate (and in some cases from the highest national health-authority); they may be granted for a defined goal, specific data type and gathering interval, and limited in time (as defined in the project's ethical protocol). In Australia, Austria, and Finland, the processing of personal data is limited to purposes of health/medical sciences and historical/statistical studies. Finland requires additional research permissions from the *Finnish Ministry of Social Affairs and Health* for projects that use personal data from more than one jurisdiction (i.e., municipality/municipality board) or from the private sector (i.e., private healthcare services or self-employed healthcare professionals). Applications are submitted for free to the *Finnish National Institute for Health and Welfare* and/or *Finnish Medicines Agency* and their review may take from three to twelve months. The *Finnish Office of the Data Protection Ombudsman* must be notified using a standardized form, if the project keeps a personal data register; transfers personal data to outside the EU or the European Economic Area; or launches an automated decision-support system. In Switzerland, furnishing ethics approvals and research permissions is amalgamated and examined by each larger healthcare provider and smaller canton's own ethics commission.

Third, *data collection* takes place for the purposes specified above. In general, the legislation in the five countries does not permit organizations to use or disclose personal data unless one or both of the following conditions applies: 1) *Data subjects* (the patients/customers and sometimes also the healthcare workers and informing clinicians) *have been consented*. 2) The *use/disclosure has been specifically authorized/required by another law*. In Austria, all data must be de-identified without delay so that the data subjects are no longer identifiable if specific project phases can be performed with indirectly personal data only. In Australia, this de-identification is encouraged. The USA allows researchers to access to certain de-identified health information and specifies this de-identification as follows: 1) removing all specific identifiers of the individuals and their relatives, employers, and household members or 2) obtaining a qualified statistical expert's documented opinion stating that the remaining risk of identifying an individual is very small. *Data gathering and de-identification*, typically by the

**Table 1. Comparison of the five countries**

|  | Australia | Austria | Finland | Switzerland | USA |
|---|---|---|---|---|---|
| **Frameworks** | Australian Government, Office of the Australian Information Commissioner Privacy Act 1988 at www.oaic.gov.au/ NHMRC National Statement on Ethical Conduct in Human Research, 2007 at www.nhmrc.gov.au/guidelines-publications/e72 | The Austrian Data Protection Commission/Authority. Austrian Federal Act concerning the Protection of Personal Data, DSG 2000 at www.dsb.gv.at/site/6274/default.aspx European Parliament, Council. Directive 95/46/EC of the European Parliament and of the Council of 24 October 1995 on the Protection of Individuals with regard to the Processing of Personal Data and on the Free Movement of such Data. http://eur-lex.europa.eu/homepage.html (EC (2012): *Proposal AND Report on Proposal for a Regulation of the European Parliament and of the Council on the Protection of Individuals with regard to the Processing of Personal Data and on the Free Movement of such Data (General Data Protection Regulation)*. Brussels, Belgium) | Statutes of Finland at www.finlex.fi: Act on Electronic Prescriptions 61/2007, Act on Private Health Care 152/1990, Act on the National Personal Data Registers for Health Care 556/1989, Act on the Status and Rights of Patients 785/1992, Act of Reading Health or Social care client information 159/2007, Criminal Code of Finland 39/1889, 940/2008, Decree of Ministry of Social Affairs and Health about Patient Records 298/2009, Health Care Law 1326/2010, Law on Medical Research 488/1999, 794/2010, and Personal Data Act 523/1999. EC Directive 95/46/EC *(and the 2012 EC proposal and report)* | National/cantonal principles relatively similar to the EC and international frameworks; see the Swiss eHealth strategy at www.e-health-suisse.ch | US Department of Health & Human Services. Health Information Privacy. The HIPAA Privacy Rule 2002. http://www.hhs.gov/hipaa/for-professionals/privacy/index.html |
| **Highest authority** | Department of Health and Ageing, National Health and Medical Research Council | Ministry of Health | Ministry of Social Affairs and Health | Ministry of Health | U.S. Department of Health & Human Services |
| **Regional differences** | Each state/territory has its own legislation. E.g., in NSW, www.legislation.nsw.gov.au: Health Records and Information Privacy Act 2002 (NSW), NSW Health Privacy Manual, Version 2, 2005, NSW Health Electronic Information Security Policy, Version 1, 2005, Privacy and Personal Information Protection Act 1998 (NSW) | (Practically) none | None | Each canton has its own legislation. E,g,, for the canton of Valais, see www.swissethics.ch/ek_detail_pages/vs_f.html | Each state/territory has its own legislation. E.g., in California the California Confidentiality of Medical Information Act, www.leginfo.ca.gov/cgi-bin/calawquery?codesection=civ |
| **Ethics approval required for** | Any health information acquisition or work with patients beyond routine care | Any work with patients beyond routine care. Exception: Non-interventional studies (NIS, i.e., studies that observe and collect data on the application of approved medicinal products, devices and procedures) are in general not subject to ethics clearance if they do not deal with identifiable biosamples or datasets. However, such projects may still be subject to "Good Scientific Practices" as established at the respective institution. As far as registries are concerned, some Ethics Committees refer to "Volume 9A of The Rules Governing Medicinal Products in the European Union", in particular to Table I.7.A that deals with post-authorization safety studies, i.e. refers to pharmacovigilance issues (ec.europa.eu/health/files/eudralex/vol-9/pdf/vol9a_09-2008_en.pd). Furthermore, NIS need to be registered with the "Bundesamt für Sicherheit im Gesundheitswesen". | Any health information acquisition or work with patients beyond routine care. Exception: Record studies without connections to invasive studies do not require a medical ethics approval. | Any health information acquisition or work with patients beyond routine care or quality control of machines | Any health information acquisition or work with patients beyond routine care or quality control |
| **Approving authorities** | The Australian Health Ethics Committee supported by territorial and site specific committees | The 27 Ethics Committees in Austria whose jurisdiction depends on where the healthcare provider involved in a clinical research project is located | The National Committee on Medical Research Ethics (TUKIJA, www.tukija.fi) and its sub-committees for the five university hospital districts. TUKIJA gives statements on studies that involve patients and that are invasive in nature (either physically or mentally) | For large hospitals (e.g., Geneva), the hospital's ethics commission. For smaller hospitals (e.g., Sion), cantonal ethics commissions | Internal Review Boards (IRBs) of institutions holding healthcare data |
| **Forms** | The NEAF National Ethics Application Form, which may need to be supplemented with territorial applications. www.neaf.gov.au | Standardized forms (e.g., Antrag auf Beurteilung eines klinischen Forschungsprojektes) | Standardized forms of TUKIJA, dependent on the project type, that may need to be supplemented with the standardized forms by the hospital districts | Each commission has its own standardized forms and structure for an application. | Each IRB has its own forms. |
| **Review times** | The committees meet every month. | Approximately three months starting with submission, the larger committees meet approximately ten times per year. | The committees meet once or twice a month. | The hospital commissions meet regularly and the cantonal commissions 6-8 times per year. | Some institutions meet every week, others fortnightly or monthly. |
| **Review costs** | 0 – over 5,000 USD, depending on commercial sponsorships of the project | 0 – over 5,000 USD, depending on project factors like commercial sponsorships and multi-centricity | 0–3,000 USD, depending on commercial sponsorships of the project | 0–1,000 USD, depending on the canton | 0 USD |
| **Data subject** | Patient/customer Healthcare workers | Patient/customer Informing physician | Patient/customer | Patient/customer | Patient |
| **Using/dis-closing identifying information for scientific, historical or statistical purposes is NOT permitted without data subjects' informed consent unless** | Specific authorization or requirement by another law: health information can be used and disclosed for health and medical research purposes in certain circumstances, where researchers are unable to seek individuals' informed consent; if it is not practicable to consent data subjects, de-identified information should be used and if also this option is unavailable, identifying information may be used if the proposed medical project has been approved by a properly constituted Human Research Ethics Committee. | Data are publicly accessible; have been lawfully collected for other research projects or other purposes; or are only indirectly personal. Other data shall only be used under special conditions, in particular if they are pursuant to specific legal provisions or with the consent of the data subject or with a permit of the Austrian Data Protection Authority. Data subjects' informed consent is needed in particular for data that have been collected in previous trials or for routine clinical treatment. Informed consent forms, which clearly state how data will be processed, are a central part of the Ethics Committee submission already. | The research cannot be carried out without data identifying the person and the consent of the data subjects cannot be obtained owing to the quantity of the data, their age or another comparable reason; the use of the personal data file is based on an appropriate research plan and a person or a group of persons responsible for the research have been designated; the personal data file is used and data are disclosed therefrom only for purposes of historical or scientific research and the procedure followed is also otherwise such that the data pertaining to a given individual are not disclosed to outsiders; and after the personal data are no longer required for the research or for the verification of the results achieved, the personal data file is destroyed or transferred into an archive, or the data in it are altered so that the data subjects can no longer be identified. | The amount of consenting work is not in relation to the amount of protection required or many of the potential subjects' might be deceased, etc. | Limited datasets in which identifiers have been removed and situations in which it is impractical to contact data subjects. |

| Information exchanges possible if | The transferring agency/organization (A/O) remains accountable for these data, unless at least one of the following three conditions applies: 1) Data subjects have been consented after being expressly advised that the consequence of providing consent is that after the transfer, the A/O will no longer accountable for the individual' personal data. 2) The A/O is required or authorized by or under law to transfer the personal data. 3) The transferring A/O reasonably believes that the data recipient is subject to a law, binding scheme, or contract which effectively upholds privacy protections that are substantially similar to these principles. | At least one of the following three conditions applies: 1) Data subjects have been consented. 2) The exchange has been authorized by the Austrian Data Protection Authority. 3) Data are only indirectly personal. | At least one of the following three conditions applies: 1) Data subjects have been consented. 2) The protection of the privacy and the rights of data subjects is guaranteed by means of contractual terms. 3) The country in question guarantees an adequate level of data protection. Additional permission are needed for data from more than one jurisdiction or from the private sector. | Specifically requested in the ethics protocol and the proper permissions have been furnished | Data users are approved by their organizational review boards |
|---|---|---|---|---|---|

healthcare service provider or an organization such as the *Australian Centre for Health Record Linkage*, may be subject to substantial fees and sometimes also delays.

Fourth, *research*, where the transferred data are used only for the purposes specified above, takes place; otherwise Steps 1-3 must be repeated. Researchers' abilities to move and combine data across geographical or jurisdictional borders is important not only for collaborative and comparative purposes but also for availability of *storage and compute cloud services* (e.g., by *Amazon*, *Dropbox*, and *Google*): the cloud may not only forbid the storage and processing of personal data in its user agreement but also use computer servers that are physically located overseas, leading to trans-border information exchanges. In general, trans-border information exchanges are permitted with data from the five countries if it has been specifically requested in the ethics protocol and the proper permissions have been obtained. For academic partners this can, in some cases, be easier to obtain than exchanges with commercial goals.

Fifth, *data destruction* is performed as specified in the ethics protocol. Typically, this means deleting all data or returning it to the jurisdiction when the research activities are finished (e.g., Finland and Switzerland). Justifying the exchange of *original data* or even *derived resources* (i.e., secondary data) is often difficult, because data subjects should have been informed about all purposes of data collection and use before data collection (e.g., all our five countries) and this information disclosure can occur only after assuring that individuals are no longer identifiable (e.g., Austria, Finland, and the USA). The requirements are particularly difficult in evolving research activities and in data mining projects, where the aim is to infer new, previously unknown information from data. For example in Austria, there is currently no generally accepted process for disseminating health data from one research project to another secondary use; processes take place – so far – on an individual and case-by-case basis and many people fear that secondary use of the national EHR data could lead to unwanted transparency and serve for unwanted control purposes [11].

## 4. DISCUSSION

The frameworks differ between the five countries and their regions, but the differences within each country are not as substantial as across countries and because most frameworks apply the principles of accountability and/or adequacy they are actually fundamentally similar. In the *accountable principle*, the original creator of the personal data register is accountable for regulatory compliance unless the accountabilities are specified separately, as in the APEC, Australian, and US frameworks, whilst in the *adequacy principle*, the subsequent information receiver must protect privacy adequately, as in the Australian and EC frameworks.

ICT can be built not only to enforce and audit compliance with all frameworks and framework updates simultaneously but also to prevent the possibility of accountability and adequacy violation in exchanging health information [12]. Building this preventative ICT that implements core privacy principles, adopts trusted network design characteristics, oversees accountability and adequacy will bolster trust in such systems and promote their adoption [13]. Such ICT for EHR exchanges between China and Japan has been introduced [14, 15]. Moreover, an access control scheme and other security designs applicable for these privacy-sensitive data have been designed for cloud computing and wireless networking [16-18]. Finally, ICT to automatically de-identify EHR text have been evaluated to reach the F1 correctness percentage from 81 to 99 in English, French, Japanese, and Swedish [19-24]. Approximately ninety per cent of the residual identifiers left behind by either these ICT or human coders can be concealed by applying additional computation methods [25].

However, similarly to our conclusion on differing frameworks, current ICT and architectures for health information exchanges are also very different when comparing between countries or jurisdictions [26]. Moreover, evolving healthcare and science (e.g., ICT development in non-research organizations or exchange of secondary data) and lessons learnt in information exchanges calls also for continuous improvements to enforcement mechanisms in the existing law and data subjects' consenting [13, 27-30]. The attitudes of 62 US patients toward electronic health information exchanges have been studied [31]. Regardless of their enthusiasm about their capacity to improve the quality and safety of healthcare through the exchanges, they are also concerned about the potential of these exchanges to result in data misuse and compromised privacy. Consequently, they want more information about data subjects' consenting.

## 5. ACKNOWLEDGMENTS

H Suominen and L Hanlen were partially funded by the Australian Government through the Department of Communications and the Australian Research Council through the Information and Communications Technology (ICT) Centre of Excellence Program (NICTA Centre of Excellence); H Müller by the EU's 7th framework program (Khresmoi project); L Ohno-Machado by the US National Institutes of Health (grants U54HL108460 and UL1RR031980) and the US Agency for Healthcare Research and Quality (grant R01HS019913); S Salanterä by the Academy of Finland (project 140323) and Tekes, the Finnish Funding Agency for Technology and Innovation (project 2227/31/2010); and G Schreier's by the EU's 7th Framework Programme (FP7 2007-2013 grant 261743). H Suominen received a travel grant by iDASH for this collaborative work.